\documentclass[conference]{IEEEtran}
\IEEEoverridecommandlockouts
\usepackage{lipsum} 
\newcommand\blfootnote[1]{%
  \begingroup
  \renewcommand\thefootnote{}\footnote{#1}%
  \addtocounter{footnote}{-1}%
  \endgroup
}
\usepackage{scrextend}
\makeatletter
\renewcommand{\footnoterule}{%
  \kern-3pt
  \hrule\@width 1.5in 
  \@height 0.4pt 
  \kern 2.6pt
}
\makeatother

\usepackage{cite}
\usepackage{booktabs}
\usepackage{amsmath,amssymb,amsfonts}
\usepackage{algorithmic}
\usepackage{graphicx}
\usepackage{textcomp}
\usepackage{xcolor}
\def\BibTeX{{\rm B\kern-.05em{\sc i\kern-.025em b}\kern-.08em
    T\kern-.1667em\lower.7ex\hbox{E}\kern-.125emX}}
\begin{document}

\title{Mitigating Category Imbalance: Fosafer System for the Multimodal Emotion and Intent Joint Understanding Challenge\\
}

\author{
\IEEEauthorblockN{Honghong Wang, Yankai Wang, Dejun Zhang, Jing Deng, Rong Zheng}
\IEEEauthorblockA{\textit{Beijing Fosafer Information Technology Co., Ltd., Beijing, China}\\
\{wanghonghong, wangyankai, zhangdejun, dengjing, zhengrong\}@fosafer.com}\\
}


\maketitle

\begin{abstract}
This paper presents Fosafer's approach to the Track 2 Mandarin in the Multimodal Emotion and Intent Joint Understanding (MEIJU) challenge, which focuses on achieving joint recognition of emotion and intent in Mandarin, despite the issue of category imbalance. To alleviate this issue, we use a variety of data augmentation techniques across text, video, and audio modalities. Additionally, we introduce the Sample-Weighted Focal Contrastive (SWFC) loss, designed to address the challenges of recognizing minority class samples and those that are semantically similar but difficult to distinguish. Moreover, we fine-tune the Hubert model to adapt the emotion and intent joint recognition. To mitigate modal competition, we introduce a modal dropout strategy. For the final predictions, a plurality voting approach is used to determine the results. The experimental results demonstrate the effectiveness of our method, which achieves the second-best performance in the Track 2 Mandarin challenge\blfootnote{This work is supported by the National Key Research and
 Development Program of China (No.2022YFF0608504)}.
\end{abstract}


\section{Introduction}
As human-computer interaction (HCI) technology rapidly advances, emotion recognition has emerged as a pivotal area of research. However, existing emotion recognition methods frequently encounter challenges in accurately capturing the user intent. Furthermore, emotion and intent recognition are often treated as distinct processes, presenting a challenge for systems to comprehensively address user needs. To address the limitation above, the Multimodal Emotion and Intent Joint Understanding Challenge (MEIJU) 2025 challenge \cite{b1} seeks to simultaneously predict both emotion and intent by integrating features from text, video, and audio modalities. 

The Track 2 Mandarin of MEIJU2025 focuses on class-imbalanced joint emotion and intention recognition.  Our strategies include addressing class imbalance through data augmentation, introducing the Sample-Weighted Focal Contrastive (SWFC) loss function \cite{b2} to improve the recognition of minority class samples, and fine-tune Hubert to extract high-quality semantic representations related to emotions and intentions. Additionally, to mitigate potential modal competition during fusion, we employ a modal dropout strategy \cite{b3}, further improving model performance.
\section{Proposed Method}
\begin{figure}[htbp]
\centerline{\includegraphics[width=0.5\textwidth]{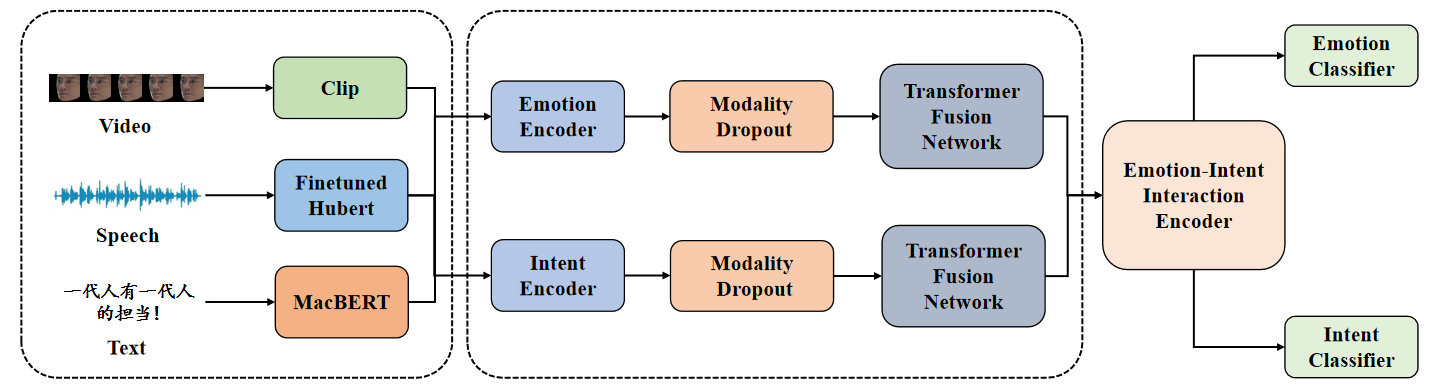}} 
\caption{Framework of our proposed method.}
\label{fig}
\end{figure}

Fig.~\ref{fig} illustrates the overall architecture of the proposed model. First, multiple single-modal feature extractors are employed to process audio, video, and text sequence data. The extracted features are then transformed into emotion and intent embeddings using dedicated encoders. Following this, a modal dropout strategy is applied before the features are integrated using a Transformer-based fusion network. The relationship between emotion and intent is learned through the emotion-intent interaction encoder. Finally, the emotion and intent classifiers predict the respective emotion and intent based on the processed features.

\subsection{Unimodal Feature Encoder}\label{AA}
Based on the findings of MERBench \cite{b4}, we select the Chinese-Hubert-large\footnote{https://huggingface.co/TencentGameMate/chinese-hubert-large}, which incorporates two classification heads. The model is then fine-tuned using the MEIJU Track 2 Mandarin data. Due to the limited scope of the fine-tuning data, the Hubert feature extractor is frozen. For visual feature extraction, we use the OpenFace toolkit\footnote{https://github.com/TadasBaltrusaitis/OpenFace} to capture facial information from the original video, followed by feature extraction using the CLIP\footnote{https://github.com/openai/CLIP}. For the text modality, recognizing the unique characteristics of Chinese language scenes, we choose the Chinese-MacBERT-large\footnote{https://huggingface.co/hfl/chinese-macbert-large} for text feature extraction.

\subsection{Data augmentation and class imbalance loss function}
In Track 2 Mandarin, we perceive that the ``acknowledging" and ``encouraging" intent labels have significantly fewer samples than the other categories. To mitigate this imbalance, we apply data augmentation techniques to all three modalities.

For the text modality, we use GPT-4 \cite{b5} to analyze and rephrase existing text, creating new samples that closely align with the original emotion and intent labels. For the video modality, new data are generated using the vidaug tool\footnote{https://github.com/okankop/vidaug}, applying random cropping and rotation to prevent the introduction of unnecessary noise. For the audio modality, we expand the samples using so-vits-svc\footnote{https://github.com/svc-develop-team/so-vits-svc}, ensuring that samples with background music are filtered out to maintain optimal audio quality.

We introduce the SWFC loss function to alleviate class imbalance. The SWFC loss extends the Focal Contrastive Loss \cite{b6} by incorporating two key components: a focus parameter and a sample weight parameter. These additions improve training optimization by differentiating between positive and negative samples. The focus parameter directs the model's attention to harder-to-classify samples, while the sample weight parameter adjusts the penalty for samples from underrepresented classes. 

\subsection{Modality Dropout Strategy}
Inspired by \cite{b3}\cite{b7}, we introduce the modal dropout strategy to mitigate competition between modalities and address modal heterogeneity, effectively. The embeddings from different modalities, generated by the emotion and intent encoder are passed through a dropout module before multimodal fusion. During training, the dropout module randomly sets the embedding of each modality to zero with a specific probability.
\section{Experiments and Results}
\subsection{Datasets and Evaluation Metric}\label{SCM}
The experiments are conducted using both the original data from Track 2 Mandarin and the augmented data. The training, validation, and test sets consist of 8331(7582 original and 749 enriched), 1083, and 2166 samples, respectively.  

Due to the category imbalance observed in Track 2, organizers use the Joint Recognition Balance Metrics (JRBM) to evaluate the model's performance in the joint emotion and intent recognition task.
\subsection{Results and Analysis}\label{SCM}
Table 1 demonstrates the result of our approach on the Track 2 Mandarin test set. Compared to the baseline system, our methods achieve a significant improvement, with the final model's JRBM score increasing by approximately 3.39\%. The table shows that our method positively impacts the recognition results. Notably, the fine-tuned Hubert model and the SWFC loss have a more substantial impact on the results.

To further optimize the prediction results, we employ a voting strategy. During training, we select multiple model checkpoints that yield the best performance on the validation set. The predictions from three models are then combined, with the final result determined by a majority vote. In cases of disagreement, the prediction from the first model is chosen. This process continues until no further improvements are observed. As a result, we achieve a JRBM score of \textbf{0.7356}, securing second place in the Track 2 Mandarin challenge and highlighting the effectiveness of our approach.

\begin{table}
\centering
\caption{Recognition results of Test set.The  baseline system adopts macbert ,CLIP and HuBERT as feature extractors while ours uses macbert ,CLIP and fine-tuned Hubert.}
\label{tab1}
\begin{tabular}{cc} 
\toprule
\textbf{System}           & \textbf{JRBM}  \\ 
\midrule
Baseline                  & 0.6680         \\
Ours                      & \textbf{0.7019}         \\
Ours w/o Data Aug         & 0.6962         \\
Ours w/o SWFC Loss        & 0.6892         \\
Ours w/o Modality Dropout & 0.6986         \\
Ours w/o Finetune HuBERT  & 0.6884         \\
\bottomrule
\end{tabular}
\end{table}

\section{Conclusion}
 In this paper, we present our solutions for the MEIJU2025 Track 2 Mandarin challenge,  in which we achieved second place. To address the issue of category imbalance, we apply data augmentation techniques to the audio, video, and text modalities. We also introduce the SWFC loss function to optimize model performance, with a focus on improving recognition for minority class samples. To mitigate the modal competition issue, we implement a modal dropout strategy. Additionally, we fine-tune the Chinese-Hubert-Large specifically for the emotion and intention joint recognition. Finally, we apply a plurality-based voting strategy to further improve the score, demonstrating the effectiveness of our approach.

\bibliographystyle{IEEEtran}

\end{document}